%% file: gencal2.tex
\documentclass[12pt]{article}

\usepackage{color}
\definecolor{brown}{rgb}{0.4,0.0,0.0} 
\definecolor{olive}{rgb}{0.0,0.4,0.0} 
\definecolor{dblue}{rgb}{0.0,0.0,0.4} 
\definecolor{purple}{rgb}{0.4,0.0,0.4} 
\definecolor{bluegreen}{rgb}{0.0,0.4,0.4} 
\definecolor{lightbrown}{rgb}{0.4,0.4,0.0} 
\definecolor{orange}{rgb}{1.0,0.65,0.0} 
\definecolor{gray}{rgb}{0.4,0.4,0.4} 

\definecolor{row1}{RGB}{255, 0, 0}
\definecolor{row2}{RGB}{0, 255, 0}
\definecolor{row3}{RGB}{0, 0, 230}
\definecolor{row4}{RGB}{255, 127, 0}
\definecolor{row5}{RGB}{139, 69, 19}
\definecolor{row6}{RGB}{127, 255, 0}
\definecolor{row7}{RGB}{0, 255, 127}
\definecolor{row8}{RGB}{127, 0, 255}
\definecolor{row9}{RGB}{0, 127, 255}
\definecolor{col1}{RGB}{140, 140, 0}
\definecolor{col2}{RGB}{144, 196, 40}
\definecolor{col3}{RGB}{0, 200, 200}
\definecolor{col4}{RGB}{200, 200, 127}
\definecolor{col5}{RGB}{200, 100, 150}
\definecolor{col6}{RGB}{127, 200, 200}
\definecolor{col7}{RGB}{100, 100, 127}
\definecolor{col8}{RGB}{100, 127, 100}
\definecolor{col9}{RGB}{127, 100, 100}
\definecolor{dia1}{RGB}{230, 230, 0}
\definecolor{dia2}{RGB}{255, 0, 255}
\definecolor{etc}{RGB}{127, 127, 127}

\title{Rational Solutions of Underdetermined Polynomial Equations}
\author{Thomas Wolf, {\tt twolf@brocku.ca}\\ 
Chimaobi Amadi, {\tt ca14se@brocku.ca} \\ \\
Department of Mathematics, Brock University\\ 
500 Glenridge Avenue, St.Catharines, \\
Ontario, Canada L2S 3A1
}
\begin{document}
%\date{January 31, 2016}
\maketitle
\begin{abstract}
In this paper we report on an application of computer algebra in which
mathematical puzzles are generated of a type that had been widely used
in mathematics contests by a large number of participants worldwide.

The algorithmic aspect of our work provides a method to compute
rational solutions of single polynomial equations that are typically
large with $10^2\ldots 10^5$ terms and that are heavily
underdetermined.  This functionality was obtained by adding modules
for a new type of splitting of equations to the existing package {\sc
  Crack} that is normally used to solve polynomial algebraic and
differential systems.
\end{abstract}

\section{Motivation} \label{motiv}
Mathematical puzzles of the form
\input{expro0.tex}
\end{center}
\rm are known for a long time. In there each different letter
represents a different digit. For example, {\tt \verb@ab@}
represents a 2-digit integer with two different digits. The challenge
is to find the value of each letter such that all 3 horizontal and 3
vertical equations are satisfied.  A newer form 
\noindent \tt \begin{center}
\input{expro1.tex}
\end{center}
\rm with two additional diagonal relations has been created and first
published by one author (TW) on the home page of Caribou Contests \cite{CCInc}
in January 2010. Since then each day a new puzzle is shown together with the
solution of the problem from the day before. Puzzles like the first one have
been used in many Caribou Contests which are run six times a year. In first
contests of the 2015/16 school year typically 16,000 students from 15
countries participated each time.

This type of problem is very popular among students. The NEAMC (North
Eastern Asian Mathematics Competition) adopted them and now holds each
year one event dedicated exclusively to such problems, taken them with
permission from the Caribou website, 2015 for the first time with
schools from 9 countries participating.

To make such puzzles even more interesting one can ask what it is that makes
puzzles popular. What wide spread puzzles like sudoku and Rubik's cube have in
common is that with the last step of their solution several conditions become
suddenly fulfilled which gives a deep gratification to the problem
solver. With this concept in mind the aim of our application is to create
puzzles with even more conditions being suddenly satisfied when the puzzle is
finally solved. In the following problem 36 conditions need to be fulfilled:
\noindent \tt \begin{center}
\Large
\input{p4485.tex}
\end{center}
\normalsize \rm
%\small
%\input{pro549.tex}
Not only the values of all 7 rows and 7 columns need to be zero, but
also the values of all 11 diagonals running from the top left to
the bottom right and the 11 diagonals running from the top right to the
bottom left need to evaluate to zero. In these enlarged puzzles the usual 
order of the operations applies ($\times, \div$ before $+, -$).

In \cite{ZW2015} a $7\times 7$ problem is shown with all 49 numbers being 
integers. To create such a challenge one 
\begin{enumerate}
\item fills the $(7+6)\times(7+6)$ matrix with randomly selected
  operators $+, -, \times, \div$, and with $7\times 7$ variables (not
  sequences of letters representing digits like in the diagram above),
\item formulates the system of $7+7+11+11=36$ polynomial equations for 
  the 49 variables,
\item computes rational solutions for this system with possibly several 
  free parameters,
\item repeats the following steps:
  \begin{itemize}
  \item replaces free parameters by random integer numbers, so that
    divisors and denominators do not become zero,
  \item encodes digits by letters,
  \item determines all solutions of the resulting puzzle
  \end{itemize}
  until the created puzzle has only one solution. 
\end{enumerate}
The following section describes the computational problem.  It is
followed by a description of the basic idea for the key computational
step, its variations and options and integration with standard
techniques.  A test to recognize the impossibility of rational
solutions early speeds up the computation.  After a description of our
implementation we report on example computations. The paper closes
with a discussion on computational aspects and a summary.

\section{The Computational Problem}
In the rest of this paper we are concerned with the generation of rational
$7\times 7$ puzzles. The system of 7+7+11+11=36 equations involves 49
unknowns, so 13 more unknowns than equations.

The case of $5\times 5$ puzzles with only two diagonal conditions is very
similar in that we have 5+5+1+1=12 conditions for 25 unknowns, i.e.\ also 13
more unknowns than equations.

The nonlinearity of the problem can be chosen in advance through the number of
multiplications and divisions that occur in the puzzle. Around five $\times$
or $\div$ operations already create challenging algebraic problems that can
not be solved in general and we do not know of computer algebra systems with
modules that could find rational solutions of large underdetermined single
equations or systems of polynomial equations.

If the puzzle contains only a limited number of multiplications and divisions
then many conditions are either completely linear or they contain at least one
linearly occuring variable, say $u_i$. In such an equation $0=A_i u_i + B_i$
the $A_i,B_i$ are polynomials in any other unknowns $u_k$ except $u_i$. In
this case progress can be made by considering the cases $A_i=0$ and $A_i \neq
0$ i.e.\ by eliminating $u_i = - B_i/A_i$.

If no single $u_i$ occurs linearly and if factorization over the rationals is
not possible then the next tool of choice are steps towards computing a
Gr\"{o}bner bases, i.e.\ reductions and the computation of S-polynomials. For
that one needs to define an ordering of variables. Lexicographical orderings
are notoriously computationally expensive. Using instead a graded
lexicographical ordering or total degree ordering keeps the total degree of
generated polynomials low but only at the prize of increasing the degree
evenly between the variables. This is expensive if there are many variables
allowing a huge number of products of powers between them, like in our
application. But even if the Gr\"{o}bner basis can be computed, then mostly
all equations are non-linear in all $u_k$ and one has not made much progress
towards rational solutions.

Often just one non-factorizable polynomial equation remains to be solved which
is non-linear for all its variables and which is just one equation so
Gr\"{o}bner basis computations do not apply. 

%But even if 2 or 3
%equations with 10,000 - 100,000 terms remain, the number of possible
%orderings to apply given that there are many unknowns, is huge and any
%one ordering that we tested leads to an explosion of the number of
%terms.

On the positive side, we have 13 more variables than equations,
i.e.\ our system is heavily underdetermined. Furthermore, we do not
need the general solution. Special rational solutions are useful for
our purpose.

\section{The Basic Idea} \label{BasicIdeas}
If the general solution is not required then additional equations can be added
if the new system can then be solved with available techniques 
in terms of rational solutions. As long as
the new system is still underdetermined, the danger of excluding all solutions
is negligible. Our strategy will be to add equations by partitioning,
i.e.\ splitting existing equations.

Let us assume a single equation is to be solved which is non-linear in
all variables $u_m$, is not factorizable over the integers and is of
degree $d_i$ in $u_i$ and therefore can be written as
\begin{equation}
0 = P(u_j) = \sum_{n=0}^{d_i} A_{in}u_i^n    \label{e1}
\end{equation}
where $A_{in}$ is the coefficient of $u_i^n$ and is a polynomial that may 
involve all unknowns $u_k$ except $u_i$.

{\bf Method 1)} The first approach is to split completely w.r.t.\ $u_i$
and solve the system
\begin{equation}
0 = A_{in}  \ \ {\rm for} \ \ n=0\ldots d_i .  \label{e2}
\end{equation}
$u_i$ becomes a free parameter in any obtained solution. 
These $d_i+1$ new equations should typically involve up to $d_i+1$
extra different unknowns to admit solutions. 

The new system (\ref{e2}) can be investigated by standard techniques.
For example, although the original equation (\ref{e1}) was non-linear
in all variables, the equations of the new system (\ref{e2}) might be
either linear in some $u_p$ or be factorizable with one factor being
linear in at least one $u_q$ and thus the new system could be solvable
simple by performing factorizations and substitutions.

{\bf Method 2)} A less restrictive system derived from partial splitting 
requires at most $d_i-1$ extra unknowns, i.e.\ 2 unknowns less:
\begin{eqnarray}
0 & = & A_{in}  \ \ \ \ \ \ \ \ {\rm for} \ \ n=2\ldots d_i  \label{e3} \\
0 & = & A_{i0} + A_{i1} u_i                       \label{e4}
\end{eqnarray}
where (\ref{e4}) can be used for a (potentially case generating)
substitution of $u_i$. The value of $u_i$ is rational as long as the
values of the other $u_k$ are rational.

{\bf Method 3)} It is even less restrictive to split (\ref{e1}) only once into
\begin{eqnarray}
0 & = & \sum_{n=2}^{d_i}A_{in}u_i^{n-2}             \label{e5} \\
0 & = & A_{i0} + A_{i1} u_i  .                    \label{e6}
\end{eqnarray}
Here $u_i$ can also be eliminated completely (in the case $A_{i1}\neq
0$). With only one extra equation, only one extra $u_k$ may be
required for (\ref{e5}), (\ref{e6}) to admit a solution. 
For $d_i=2$ methods 2) and 3) are identical.

An additional benefit of methods 2), 3) compared to method 1) is that $u_i$
can be replaced also in all other equations that may have to be solved apart
from (\ref{e1}).

\section{Variations and Options} \label{VandO}
The three methods of equation splitting have several parameters and
can be combined with standard solving techniques in different ways 
and with themselves recursively. For their optimal use one may 
have to strike a balance between
\begin{itemize} \itemsep -3pt
\item finding many different solutions,
\item finding solutions involving a maximum number of free parametric variables,
\item managing computation times,
\item managing computer memory,
\end{itemize}
all depending on the nature of the application. For example, one may need to
solve one specific system and put maximal effort into this one investigation, 
or as in our
case when generating puzzles, one may have many similar systems and could
collect rational solutions for any one system and rather start investigating a
new system before equations get too large.

The following are questions and answers concerning the three methods. Comments
about their integration with other computational steps will be made in the
next section.
\begin{itemize}
\item[A)] {\em Does the order of splittings matter?} \\ The operations to
  split w.r.t.\ different variables commute, i.e.\ it does not matter whether to
  split first w.r.t.\ $u_i$ and then $u_j$ or in the opposite order as long as
  only splittings occur. The situation is different when between splittings
  also factorizations and substitutions are performed. Then the order of
  splittings is relevant. The consequence is that if it is critical to find
  the maximal number of rational solutions, then one would have to consider
  all possible orders of splittings and has to perform all possible
  factorizations and substitutions in-between.

  In our case having, for example, 3 equations left to solve after all
  substitutions and factorizations have been done, each involving 49 - 33 = 16
  variables (from the 36 original equations, 33 have been used for
  substitutions so that 3 equations remain) then $3\times 16$ different
  partial splittings can be done. These are too many to consider all
  permutations between them. Even to try all of them once as first splitting 
  would be cumbersome. Therefore heuristics which splittings to try or which 
  to try first are useful and are discussed next.
  %  The situation would be better if Gr\"{o}bner base computations would be
  %  feasible but they proved to be 
\item[B)] {\em If one has a system of equations, which equation shall be
  split with respect to which variable first/next?} \\ 
  The following heuristics are applied in this order in automatic runs.\vspace{-5pt}
  \begin{itemize}
  \item The highest priority is to split an equation with the fewest number of
    variables for the following reason.

    If a system includes a (polynomial) equation involving only one variable
    then the variable can be eliminated using this equation or no rational
    solutions exist (see section \ref{RatTest} below). Equations with 2
    variables are useful because splitting them gives one or more equations
    with only one variable which guarantee quick progress.

    Therefore equations with fewer variables lead quicker to progress through
    substitution or recognizing the non-existence of rational solutions.
  \item From the equations with the same lowest number of variables the
    advantage to split the shortest equation is that splitting creates even
    shorter equations which have the potential to shorten other long equations
    when substitutions are performed.
  \item $d_i$ should be low to avoid imposing many restrictions in splitting
    methods 1) and 2) or avoid size explosion in method 3) when $u_i$ is
    substituted in higher powers of $u_i$.
  \item $A_{i0}$ and especially $A_{i1}$ should be short to avoid length
    explosion arising from the substitution $u_i = - A_{i0}/A_{i1}$ of methods
    2) and 3).  Especially the size of $A_{i1}$ is crucial because after
    substitution all equations are written as polynomials, i.e.\ all
    previously $u_i$ independent terms will be multiplied with $A_{i1}$ to the
    power of the maximum degree of $u_i$.
   \item $A_{i0}, A_{i1}$ should involve as few unknowns as possible. After a
     substitution of $u_i$ the degree of all involved variables will typically
     increase in other equations, making later partial splittings and
     substitutions of these variables more costly.  \vspace{-5pt}
\end{itemize} 
\item[C)] {\em Which of the three splitting methods shall be used and when?}\\
  The answer depends partially on the degree of underdetermination. The more
  unknowns are available the more restrictive can be the splitting as
  performed in method 1. On the other hand, if the most general solution is to
  be found then method 3) is to be preferred.

  A disadvantage of method 3) is that the substitution of $u_i$ in the other
  equation (\ref{e5}) will most likely increase the degree of the other
  unknowns in (\ref{e6}) with the consequence that unknowns which occur
  linearly in (\ref{e3}) and allow the complete solution of all equations
  (\ref{e3}) might not occur linearly in (\ref{e5}) and thus not allow the
  solution of (\ref{e5}) purely by substitution. On the other hand, to solve
  (\ref{e5}) by substitution it is enough if only one unknown occurs linearly
  in it whereas to solve (\ref{e3}) $d_i-1$ unknowns need to occur linearly
  in a way that this system can be solved.
\end{itemize}
The partial splittings introduced in section \ref{BasicIdeas}
are useless on their own. They need to be incorporated into a solver for
polynomial systems.

\section{Integration with Standard Techniques}
Before discussing the options we want to describe the program 
environment to which one module for testing the rationality of solutions 
and 2 modules for partial separation (splitting) have
been added.
\subsection{The Package {\sc Crack} for Working with Systems of Equations}
The computer algebra package {\sc Crack} \cite{TW2004}, \cite{TWonline}, is
written in {\sc Reduce} \cite{REDUCE}. It is a tool to investigate and solve
systems of algebraic and differential equations. The following features are
relevant for the successful integration of partial splitting methods.
\begin{itemize}
\item A system of equations to be solved is always considered together with a
  set of inequalities including OR-inequalities (lists of expressions of which
  at least one needs to be non-zero). Inequalities are actively collected,
  simplified and used to simplify equations and to avoid case distinctions.
\item The computation investigates all cases and (sub-)$^n$ cases that
  come up depending on the steps that are performed, like
  factorizations, case generating substitutions and adhoc case
  distinctions motivated by known solutions.
\item About 50 modules can be applied with different parameters giving
  over 90 different calls ranging from substitutions, reductions, algebraic
  combinations for the shortening of equations, integrations and 
  separations to applying the external packages Singular or DiffAlg to
  the whole system. In addition there is a large number of diagnostics
  commands available.
\item A solution strategy is composed through a list of numbers ({\tt
  proc\_list}), each number representing a module. The modules are
  tried in the order they occur in the {\tt proc\_list} until one
  module is successful and then execution starts again at the beginning
  of the list. The list can be modified interactively but also through
  modules themselves and can easily be adapted to all types of problems:
  algebraic/differential, linear/nonlinear, medium/large.
\item Equations are associated with property lists that also include 
  results of earlier investigations to avoid duplicate checks.
\item Most importantly, the program can be run fully automatically, semi
  automatically and fully interactively by allowing not only which step is to
  be done but also with which parameters and which equation(s).
\end{itemize}

\subsection{A Rationality Test} \label{RatTest}
Because the whole computation consists of a large tree of cases and
(sub-)$^n$ cases resulting from factorizations of equations and case
generating substitutions, the finding of non-rational values of at least
one variable in the solution would allow to stop the investigation of the 
current (sub-)case and to proceed with another (sub-)case. The following
simple test of detecting non-rationality is implemented in a new module 
{\bf 89}.

If the system involves an equation in only one variable then this variable can
either be eliminated using this equation or the system has no rational solution.
If the equation is linear then it is used for substitution. If it is quadratic
then if roots are rational then 2 cases are considered and substitutions are
made otherwise no rational solutions exist. If the degree is higher than 2
then either the equation is factorizable or it has no rational solutions and
computation also stops.
 
This test could be extended using the Hasse Principle to decide on the
non-existence of rational solutions for polynomials of degree 2 in an
arbitrary number of variables. 

\subsection{Our Implementation} \label{m89} 
To extend the package {\sc Crack} with partial separations we added 3 new 
modules.

\begin{itemize}
\item[\# {\bf 89}] performs a rationality test as described above. If the
  system contains a non-linear equation in only one variable then
  \begin{itemize}
  \item if factorizability of this equation has not been completely checked
    yet then the factorization test of this equation is put on the {\tt
      to\_do\_list}, else
  \item if factorizability is know then the splitting into 2 cases of one of
    the factors either being zero or being non-zero is put on the 
    {\tt to\_do\_list}, else
  \item the equation is not factorizable, investigation of the current 
    case is terminated. \vspace{-3pt}
  \end{itemize}
  Any entry in the {\tt to\_do\_list} is executed by the first module of 
  the {\tt proc\_list}.

\item[\# {\bf 90}] implements method 2) as described in section
  \ref{BasicIdeas}.  It is only applied if none of the equations is linear in
  any $u_j$.  This is guaranteed in automatic mode because in {\tt proc\_list}
  it is placed after case-generating substitutions (module {\bf 21}).  In
  interactive mode module {\bf 90} issues a hint if substitution is possible.

  If no substitution is possible then all pairs $(P,u_i)$ of equations $0=P$
  and variables $u_i$ appearing in $P$ are discarded if \vspace{-5pt}
  \begin{itemize}
  \item any one coefficient $A_{in}$ of $u_i^n,\ n>1$ can be shown to be
    non-zero based on the inequalities that are known, or
  \item if $A_{in}$ and in the case that $A_{in}$ is factorizable also all 
    its factors are nonlinear in all their $u_k$. \vspace{-3pt}
  \end{itemize}
  For the remaining pairs all those which allow the solution of the system
  (\ref{e3}), (\ref{e4}) by the {\sc Reduce} {\tt solve} command are listed so
  that in interactive mode the user can select a pair $(P,u_i)$ and in automatic
  (batch-)mode a pair from the shortest equation $P$ is selected. 
  The corresponding equations (\ref{e3}), (\ref{e4}) are added to the current
  system of equations and execution proceeds with the start of the 
  {\tt proc\_list}. 

\item[\# {\bf 91}] implements method 3). This module is also only executed if
  no equation is linear in any one $u_i$. In addition only those pairs
  $(P,u_i)$ are considered for which is $A_{i1}\neq 0$ so that $u_i$ can be
  eliminated. From all those pairs the user can select a pair in interactive
  mode and in automatic mode one pair is selected according to weights
  discussed under {\bf B)} in section \ref{VandO}. For the selected pair
  $(P,u_i)$ an induced case splitting into the two cases $A_{i0}+u_iA_{i1} =
  0$ and $A_{i0}+u_iA_{i1} \neq 0$ is put on the {\tt to\_do\_list} to be
  executed next. In this way, if both cases are considered, no solution is
  lost. In practise the case $A_{i0}+u_iA_{i1} \neq 0$ will hardly ever be
  tried because of the many further case distinctions that would be necessary
  because $A_{i0}+u_iA_{i1} \neq 0$ itself does not provide any
  simplification.  In the case $A_{i0}+u_iA_{i1} = 0$ the other equations
  (\ref{e5}) are derived automatically later when $A_{i1}=0$ or $u_i = -
  A_{i0}/A_{i1}$ are considered.

\end{itemize}

The {\tt proc\_list} used for the computations in section \ref{ExComp} is
(1 89 20 77 47 21 38) where the numbers encode the following
modules which are continuously tried in this order. They perform, if possible,
\begin{itemize} \itemsep -3pt
\item[{\bf 1}] any urgent steps listed on the {\tt to\_do\_list} which
  initially is empty but can be filled through other modules (e.g. {\bf 89})
  during the computation. Entries on this list consist of a module number and
  a list of equations that the module is to be applied to.
\item[{\bf 89}] the rationality test described under \ref{RatTest}
\item[{\bf 20}] a substitution that does not generate a case distinction
\item[{\bf 77}] the factorization of any one equation
\item[{\bf 47}] the start of a case distinction of an equation that is
  factorized
\item[{\bf 21}] a case-generating substitution
\item[{\bf 38}] exit of automatic (batch-) mode and entering interactive mode
  because the earlier modules in this list can not make progress.
\end{itemize}
When execution is stopped because {\bf 38} was reached then one can save the
worksheet and try interactively, for example, these modules
\begin{itemize} \itemsep -3pt
\item[{\bf 90}] the partial splitting method 2) described above
\item[{\bf 91}] the partial splitting method 3) described above
\item[{\bf 27}] a polynomial reduction step according to some
  predefined ordering
\item[{\bf 30}] a Gr\"{o}bner step according to some predefined ordering
\item[{\bf 59}] the system including a monomial ordering is exported and the
  package {\sc Singular} \cite{DGPS} is called to compute a Gr\"{o}bner basis
  in a prescribed time limit which. In case of success the Gr\"{o}bner basis
  is read back into {\sc Crack} for further solution.
\end{itemize}
The calculation described in section \ref{ExComp} below runs automatically
as long as at least one module in {\tt proc\_list} can be applied. When the
computation stops, module {\bf 90} is executed interactively by simply typing
90 $<$Enter$>$. Afterwards the computation continues fully automatically based
on {\bf proc\_list} until the next time that module {\bf 90} is called
interactively. The whole computation could be run fully automatically with
the {\tt proc\_list} (1 89 20 77 47 21 90). The only reason that we execute
module {\bf 90} manually is to record the number of equations still to be solved and
their size and to report these data in the following section to give the
reader an impression of the tkind of systems to which module {\bf 90} was
applied.

Module {\bf 90} is programmed to select and perform one partial splitting and
not to start a large case distinction and finally trto y all possible splittings 
as it would be done if module {\bf 91} would be applied.

The reason that we use module {\bf 90} and not {\bf 91} is that the whole tree
of all possible cases of all splittings of all equations w.r.t.\ all $u_i$ and
that recursively would be far too large. Another reason is that we do not need
too many solutions with the same operator setting like (\ref{OpSet}) but we
rather want to repeat computations with other random operator settings to
generate differently looking puzzles.

An optimal {\tt proc\_list} is found by following general principles and using
experience.  The list also dependents on the size and nature of the
computational problem. Interactive experiments about the effect of different
modules at different stages of the solution process help also to improve the
list and to adapt it to a specific computational problem.

\section{Example Computations} \label{ExComp}
The following operator setting involves 3 multiplications and 2 divisions.
\large
\begin{equation} \label{OpSet}
\begin{array}{ccccccccccccc}
u_{1} & + &u_{2} & + &u_{3} & - &u_{4} & - &u_{5} &\times&u_{6}&-&u_{7}  \\
    - & - &  /   & - &  -   & - &  -   & - &  +   & + &  -   & + &  +    \\
u_{8} & - &u_{9} & + &u_{10}& + &u_{11}& - &u_{12}& - &u_{13}& - &u_{14} \\
    - & + &  -   & - &  -   & + &  -   & - &  -   & + &  -   & - &  +    \\
u_{15}& + &u_{16}& + &u_{17}& - &u_{18}& - &u_{19}& - &u_{20}& + &u_{21} \\
    + & + &  -   & - &  -   & + &\times& + &  +   & - &  +   & - &  +    \\
u_{22}& + &u_{23}& - &u_{24}& + &u_{25}& + &u_{26}& - &u_{27}& - &u_{28} \\
    - & - &  -   & - &  +   & + &  -   & + &  +   & - &  -   & + &  +    \\
u_{29}& - &u_{30}& - &u_{31}& + &u_{32}& + &u_{33}& - &u_{34}&\times&u_{35} \\
    + & + &  +   & - &  -   & + &  +   & - &  /   & - &  +   & + &  -    \\
u_{36}& + &u_{37}& - &u_{38}& - &u_{39}& - &u_{40}& + &u_{41}& + &u_{42} \\
    + & - &  +   & - &  -   & - &  -   & - &  -   & + &  -   & - &  +    \\
u_{43}& + &u_{44}& + &u_{45}& - &u_{46}& - &u_{47}& - &u_{48}& + &u_{49} 
\end{array}
\end{equation}
\normalsize
Using the above {\tt proc\_list} the modules 
{\bf 1, 89, 20, 77, 47, 21, 38} are tried 
in this order where module {\bf 47} starts a case distinction 
if an equation is known to be factorizable (found before by module {\bf 77})
and where module {\bf 21} starts a case
distinctions if a case generating substitution
is possible. When module {\bf 38} is executed and automatic execution
stops the partial splitting module {\bf 90} is used
for the first time in {\em case 1.1.2.2} which has one equation with 1027 terms 
in 12 variables left to be solved. Module {\bf 90} replaces it with
one equation with 1014 terms and one factorizable equation with 7 terms.
No rational solution results from this splitting

Later in {\em case 1.2} one equation with 1268 terms and 13 variables remains.
Module {\bf 90} replaces it with one equation with 1253 terms and one
factorizable equation with 15 terms. This system has 4 rational solutions
resulting in cases: \vspace{-6pt}
\begin{itemize} \itemsep -3pt
\item {\em 1.2.1.2.1.2} with 10 free parameters and 3814 terms
in numerators and denominators of the 49-10=39 expressions for the 39 computed
variables. % pz-7x7-3m-2d-777193-1   ./7x7/p2230.tex

\item {\em 1.2.1.2.2} with 11 free parameters and 27032 terms
in the expressions for the 38 computed variables. % pz-7x7-3m-2d-777193-2

\item {\em 1.2.2.2.1.2.2} with 10 free parameters and 3973 terms
in expressions for the 39 computed variables. % pz-7x7-3m-2d-777193-3

\item {\em 1.2.2.2.2} with 11 free parameters and 3973 terms
in expressions for the 38 computed variables. % pz-7x7-3m-2d-777193-4
\end{itemize}

The next time module {\bf 90} is applied is in {\em case 2.1.2} having one
equation with 605 terms and 13 variables left to be solved. This equation is
replaced by one equation with 599 terms and one factorizable equation in 5
terms. This system has no rational solution.

Afterwards in {\em case 2.2} one equation with 118879 terms in 14 variables
remains. Module {\bf 90} is applied and replaces this equation by 6 equations
with 105, 649, 2922, 10816, 35007, 69380 terms. When a substitution is
performed later, the system becomes too large to continue the computation
on the 32 GB 64 Bit linux machine running the computer algebra system {\sc Reduce}. 

The $7\times 7$ puzzle in section \ref{motiv} is obtained when in the second 
of the above four solutions the 11 parameters are replaced by some small 
integers and afterwards digits are replaced by letters. It is verified 
that this puzzle has a unique solution.

%%\input{p2230.tex}
%\Large
%\input{p4485.tex}
%\normalsize \rm

%\section{Possible Future Extensions}
%\section{Discussion}
%\subsection{Comments on the Calculation}

\section{Comments on Solutions to the Application}
A key parameter in the generation of $7\times 7$ puzzles is the number of
multiplications and divisions. The higher their number is, the more nonlinear
is the algebraic system, the more terms are involved in rational solutions and
the larger are the numbers in the puzzle.

To increase the fun part in solving large puzzles and to avoid the tedious
part of replacing a letter by a digit everywhere and checking the numerical
correctness of rows, columns and diagonals an interactive puzzle solving page
\cite{MH2015} was created. On this page (at the bottom of $7\times 7$) 
the manual solution of the above puzzle is shown.

To generate $7\times 7$ {\em integer} puzzles instead of rational puzzles one
could impose extra conditions $A_{i1}=1$ and pay with the excess of available
variables.

\section{Summary}
Apart from progress in the creation of highly rewarding mathematical puzzles a
simple method with three variations has been suggested to find rational
solutions of heavily underdetermined single equations or systems of polynomial
equations. These methods separate equations partially.  An important
ingredient to a successful application is the integration with other
conventional methods for solving polynomial systems. Such an integration with
the package {\sc Crack} was easily possible due to the modular and interactive
nature of {\sc Crack}. An example computation has been described together with
a mathematical puzzle that has been generated.

\end{document}

%% file: expro0.tex
\noindent \tt \begin{center}
\verb@ ab @\textcolor{red}{$\!\times\!$}\verb@ cd @\textcolor{red}{=}\verb@ ebaf @\\
\verb@  @\textcolor{green}{+}\verb@    @\textcolor{blue}{$\!\div\!$}\verb@     @\textcolor{cyan}{-}\verb@  @\\
\verb@egf @\textcolor{magenta}{$\!\times\!$}\verb@  h @\textcolor{magenta}{=}\verb@  hdc @\\
\verb@  @\textcolor{green}{=}\verb@    @\textcolor{blue}{=}\verb@     @\textcolor{cyan}{=}\verb@  @\\
\verb@ejc @\textcolor{dblue}{$\!\times\!$}\verb@  k @\textcolor{dblue}{=}\verb@  bfj @\\
\verb@@

%% file: expro1.tex
\verb@ @\\
\verb@ ab @\textcolor{red}{$\!\times\!$}\verb@ cd @\textcolor{red}{=}\verb@ ebaf @\\
\verb@  @\textcolor{green}{+}\verb@ @\textcolor{lightbrown}{$\!\times\!$}\verb@  @\textcolor{blue}{$\!\div\!$}\verb@ @\textcolor{orange}{$\!\div\!$}\verb@    @\textcolor{cyan}{-}\verb@ @\\
\verb@egf @\textcolor{magenta}{$\!\times\!$}\verb@  h @\textcolor{magenta}{=}\verb@  hdc @\\
\verb@  @\textcolor{green}{=}\verb@ @\textcolor{orange}{=}\verb@  @\textcolor{blue}{=}\verb@ @\textcolor{lightbrown}{=}\verb@    @\textcolor{cyan}{=}\verb@ @\\
\verb@ejc @\textcolor{dblue}{$\!\times\!$}\verb@  k @\textcolor{dblue}{=}\verb@  bfj @\\
\verb@@

%% file: p4485.tex
\[\begin{array}{ccccccccccccc}
 \frac{cgj}{af}  & \textcolor{row1}{ + } &  \frac{-ii}{j}  & \textcolor{row1}{ + } &  \frac{-edhd}{ahf}  & \textcolor{row1}{ - } & i & \textcolor{row1}{ - } & f & \textcolor{row1}{ \times } & e & \textcolor{row1}{ - } &  \frac{-fca}{eb}  \\ 
\textcolor{col1}{ - } & \textcolor{dia1}{ - } & \textcolor{col2}{ \div } & \textcolor{etc}{ - } & \textcolor{col3}{ - } & \textcolor{etc}{ - } & \textcolor{col4}{ - } & \textcolor{etc}{ - } & \textcolor{col5}{ + } & \textcolor{etc}{ + } & \textcolor{col6}{ - } & \textcolor{dia2}{ + } & \textcolor{col7}{ + } \\ 
 \frac{-ii}{j}  & \textcolor{row2}{ - } & h & \textcolor{row2}{ + } & i & \textcolor{row2}{ + } & aa & \textcolor{row2}{ - } & c & \textcolor{row2}{ - } &  \frac{hi}{j}  & \textcolor{row2}{ - } & -e \\ 
\textcolor{col1}{ - } & \textcolor{etc}{ + } & \textcolor{col2}{ - } & \textcolor{dia1}{ - } & \textcolor{col3}{ - } & \textcolor{etc}{ + } & \textcolor{col4}{ - } & \textcolor{etc}{ - } & \textcolor{col5}{ - } & \textcolor{dia2}{ + } & \textcolor{col6}{ - } & \textcolor{etc}{ - } & \textcolor{col7}{ + } \\ 
 \frac{hgca}{ahf}  & \textcolor{row3}{ + } & i & \textcolor{row3}{ + } & f & \textcolor{row3}{ - } &  \frac{a}{f}  & \textcolor{row3}{ - } &  \frac{cbjj}{ahf}  & \textcolor{row3}{ - } & j & \textcolor{row3}{ + } &  \frac{ih}{j}  \\ 
\textcolor{col1}{ + } & \textcolor{etc}{ + } & \textcolor{col2}{ - } & \textcolor{etc}{ - } & \textcolor{col3}{ - } & \textcolor{dia1}{ + } & \textcolor{col4}{ \times } & \textcolor{dia2}{ + } & \textcolor{col5}{ + } & \textcolor{etc}{ - } & \textcolor{col6}{ + } & \textcolor{etc}{ - } & \textcolor{col7}{ + } \\ 
i & \textcolor{row4}{ + } &  \frac{cfih}{ahf}  & \textcolor{row4}{ - } &  \frac{eij}{ahf}  & \textcolor{row4}{ + } &  \frac{-hhed}{eb}  & \textcolor{row4}{ + } &  \frac{ihab}{ebg}  & \textcolor{row4}{ - } &  \frac{-hfca}{fc}  & \textcolor{row4}{ - } & ah \\ 
\textcolor{col1}{ - } & \textcolor{etc}{ - } & \textcolor{col2}{ - } & \textcolor{etc}{ - } & \textcolor{col3}{ + } & \textcolor{dia2}{ + } & \textcolor{col4}{ - } & \textcolor{dia1}{ + } & \textcolor{col5}{ + } & \textcolor{etc}{ - } & \textcolor{col6}{ - } & \textcolor{etc}{ + } & \textcolor{col7}{ + } \\ 
 \frac{-cbai}{ahf}  & \textcolor{row5}{ - } &  \frac{dfeb}{ahf}  & \textcolor{row5}{ - } &  \frac{bjei}{ahf}  & \textcolor{row5}{ + } &  \frac{caeh}{fc}  & \textcolor{row5}{ + } &  \frac{eadc}{cf}  & \textcolor{row5}{ - } &  \frac{-cbcb}{ahf}  & \textcolor{row5}{ \times } &  \frac{a}{e}  \\ 
\textcolor{col1}{ + } & \textcolor{etc}{ + } & \textcolor{col2}{ + } & \textcolor{dia2}{ - } & \textcolor{col3}{ - } & \textcolor{etc}{ + } & \textcolor{col4}{ + } & \textcolor{etc}{ - } & \textcolor{col5}{ \div } & \textcolor{dia1}{ - } & \textcolor{col6}{ + } & \textcolor{etc}{ + } & \textcolor{col7}{ - } \\ 
 \frac{agja}{eb}  & \textcolor{row6}{ + } &  \frac{achf}{eb}  & \textcolor{row6}{ - } &  \frac{bja}{ahf}  & \textcolor{row6}{ - } &  \frac{eejb}{ebg}  & \textcolor{row6}{ - } &  \frac{agie}{cf}  & \textcolor{row6}{ + } &  \frac{-hgc}{cf}  & \textcolor{row6}{ + } &  \frac{-iba}{ai}  \\ 
\textcolor{col1}{ + } & \textcolor{dia2}{ - } & \textcolor{col2}{ + } & \textcolor{etc}{ - } & \textcolor{col3}{ - } & \textcolor{etc}{ - } & \textcolor{col4}{ - } & \textcolor{etc}{ - } & \textcolor{col5}{ - } & \textcolor{etc}{ + } & \textcolor{col6}{ - } & \textcolor{dia1}{ - } & \textcolor{col7}{ + } \\ 
 \frac{-aheah}{ahf}  & \textcolor{row7}{ + } &  \frac{agja}{eb}  & \textcolor{row7}{ + } &  \frac{ja}{f}  & \textcolor{row7}{ - } &  \frac{-dehi}{ahf}  & \textcolor{row7}{ - } &  \frac{fdh}{jg}  & \textcolor{row7}{ - } &  \frac{-iba}{ai}  & \textcolor{row7}{ + } &  \frac{-ahef}{eb} 
\end{array} \]